\documentclass[preprint, aip, jcp]{revtex4-1}
\usepackage{graphicx}

\begin{document}
\title{Modification of the Gay-Berne potential for improved accuracy and speed}
\author{Rasmus A. X. Persson}
\email{rasmus.persson@chem.gu.se}
\affiliation{Department of Chemistry \& Molecular Biology, University of
Gothenburg, SE-412 96 G\"oteborg, Sweden}

\begin{abstract}
A modification of the Gay-Berne potential is proposed which is about 10\% to
20\% more speed efficient (that is, the original potential runs 15\% to 25\%
slower, depending on architecture) and statistically more accurate in
reproducing the energy of interaction of two linear Lennard-Jones tetratomics
when averaged over all orientations. For the special cases of end-to-end and
side-by-side configurations, the new potential is equivalent to the
Gay-Berne one.
\end{abstract} 

\maketitle

The Gay-Berne potential \cite{gay81} is a coarse-grained potential designed
specifically to reproduce the energy of interaction between two linear,
tetratomic Lennard-Jones molecules. As such, it allows the approximate
treatment of a linear, rigid polyatomic segment by use of a single interaction
site, reducing the amount of looping -- and hence the calculational effort --
drastically when calculating energies, forces and torques in molecular
mechanics. The Gay-Berne potential is not the only single-site potential for
linear molecules or molecular moieties lacking substantial electric multipole
momoments, \cite{corner48, kihara53, kohin60, buckingham67, berne72, koide74,
walmsley77, vesely06} but it is arguably the most widely used. In
this Note, we present a variation of the Gay-Berne potential which undeniably
improves even further on the computational economy and also on the accuracy.
For ease of reference and of comparison, the original functions are also
reproduced.

\begin{figure}
\includegraphics{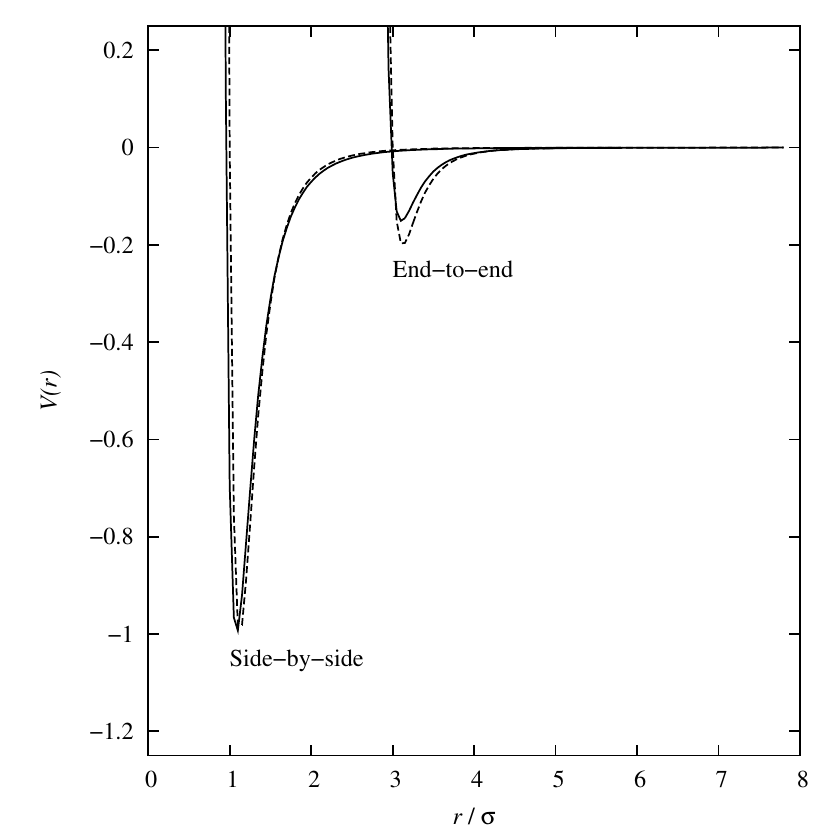}
\caption{Energy of interaction of two linear, tetratomic Lennard-Jones
molecules of bond length $2\sigma/3$ (full line) and their Gay-Berne
approximation (dashed line) with $l = 3$ and $d = 5$. The binding energy of
the side-by-side configuration is set to unity in both cases. The unit of
length is $\sigma$ of the Lennard-Jones interaction.}
\label{fig:compar1}
\end{figure}

The Gay-Berne potential, for intermolecular distance $r$, is given in eq.
(\ref{eq:gb}).
\begin{equation}
\label{eq:gb}
V(\widehat u_1, \widehat u_2, \widehat r) = \epsilon(\widehat u_1, \widehat
u_2, \widehat r) \left [ \left (\frac {\sigma_0} {r - \sigma(\widehat u_1,
\widehat u_2, \widehat r) + \sigma_0} \right )^{12} - \left (\frac {\sigma_0}
{r - \sigma(\widehat u_1, \widehat u_2, \widehat r) + \sigma_0} \right )^6
\right ].
\end{equation}
In this equation the $\epsilon(\widehat u_1, \widehat u_2, \widehat r)$ and
$\sigma(\widehat u_1, \widehat u_2, \widehat r)$ are auxiliary functions that
depend both on the molecular orientations, expressed through the unit vectors
$\widehat u_1$ and $\widehat u_2$, and on the orientation of the
intermolecular vector, expressed by $\widehat r$. Gay and Berne\cite{gay81}
provide the following forms for these functions:
\begin{eqnarray}
\label{eq:gay1}
\sigma(\widehat u_1, \widehat u_2, \widehat r) & = & \sigma_0 \left (1 - \frac
{\chi} 2 \left \{ \frac {(\widehat r \cdot \widehat u_1 + \widehat r \cdot
\widehat u_2)^2} {1 + \chi(\widehat u_1 \cdot \widehat u_2)} + \frac {(\widehat
r \cdot \widehat u_1 - \widehat r \cdot \widehat u_2)^2} {1 - \chi(\widehat u_1
\cdot \widehat u_2)} \right \} \right )^{-\frac 1 2} \\
\label{eq:gay2}
\epsilon(\widehat u_1, \widehat u_2, \widehat r) & = & \epsilon_0 \left [1 -
\chi^2 (\widehat u_1 \cdot \widehat u_2)^2 \right ]^{-\frac 1 2} \left \{1 -
\frac {\chi'} {2} \left [ \frac {(\widehat r \cdot \widehat u_1 + \widehat r
\cdot \widehat u_2)^2} {1 + \chi'(\widehat u_1 \cdot \widehat u_2)} + \frac
{(\widehat r \cdot \widehat u_1 - \widehat r \cdot \widehat u_2)^2} {1 -
\chi'(\widehat u_1 \cdot \widehat u_2)} \right ] \right \}^2
\end{eqnarray}
where $\sigma_0$ and $\epsilon_0$ are constants, and $\chi$ and $\chi'$ relate
to the anisotropy of the interaction and are defined by $\chi = (l^2 - 1) /
(l^2 + 1)$ and $\chi' = (\sqrt{d} - 1) / (\sqrt{d} + 1)$, where $l$ is the
steric length-to-breadth ratio (the axial ratio) and $d$ is the ratio of the
binding energy in the side-by-side configuration to that of the end-to-end
configuration. A comparison of the energy of interaction between the model
system it was designed to replicate, and the Gay-Berne potential itself is
shown in Figure \ref{fig:compar1}, using the original parameter set advocated
by Gay and Berne themselves. We shall keep the general form of eq.
(\ref{eq:gb}), but replace the auxiliary functions $\sigma$ and $\epsilon$ by
more efficient alternatives.

The choices for $\sigma$ and $\epsilon$ that we make are inspired by Ref.
\onlinecite{persson11c}, in which the equipotential surface of the
Lennard-Jones potential was redefined to that of an ellipsoid. There,
the philosophy of Berne and Pechukas \cite{berne72} and of Kihara
\cite{kihara53} -- that the anisotropic
intermolecular potential is obtained by replacing the distance scale by an
anisotropic function -- was strictly adhered to. In abandoning that principle
for the prescription of Gay and Berne, \cite{gay81} so that we {\em displace}
rather than {\em dilate} the interaction potential, using the anisotropic
$\sigma$ function of Ref. \onlinecite{persson11c} and its close analog for the
$\epsilon$ function in eq (\ref{eq:gb}), we arrive at the modified Gay-Berne
potential. To avoid a notational cluttering and as confusion is unlikely to
arise, we do not distinguish these new functions through notation from their
Gay-Berne counterparts. The new angular functions are represented by
\begin{eqnarray}
\label{eq:mod1}
\sigma(\widehat u_1, \widehat u_2, \widehat r) & = & \sigma_0 \left [1 + \frac
{l - 1} {2} (|\widehat r \cdot \widehat u_1| + |\widehat r \cdot \widehat u_2|)
\right ] \\
\label{eq:mod2}
\epsilon(\widehat u_1, \widehat u_2, \widehat r) & = & \epsilon_0 \left [1 +
\frac {d^{-1} - 1} {2} (|\widehat r \cdot \widehat u_1| + |\widehat r \cdot
\widehat u_2|) \right ]
\end{eqnarray}
These new functions give the same dashed curves in Figure \ref{fig:compar1} as
the original ones. Important as such agreement might be, those two geometries
only represent a subset of all possibilities; a more thorough test of the model
is to calculate its average error over all possible configurations. The average
absolute error, defined by $\langle |E_\mathrm{exact} / E_\mathrm{model} - 1|
\rangle$ where the angle brackets denote an unweighted angular average, is
shown in Table \ref{tab:compar2} for both the original and the modified
Gay-Berne potential as a function of intermolecular separation.
$E_\mathrm{exact}$ is the energy of the dimer of the linear tetratomic, as used
in the original paper, \cite{gay81} and $E_\mathrm{model}$ is the energy of
interaction of the Gay-Berne potential, original or modified, in the same
geometry. Both energies are normalized by their respective side-by-side binding
energy. It is worth pointing out that this comparison is made with the original
Gay-Berne parameter set, optimized for this situation, and that further
improvement is perhaps possible with adjustments. It is a chief strength of the
modification that it does not need reparametrization to improve upon the
original.

\begin{table}[h]
\caption{Angular-averaged unsigned errors for selected intermolecular distances
$r$ for the original and modified Gay-Berne potential with respect to the
linear Lennard-Jones tetratomic, obtained by unbiased Monte Carlo sampling over
$10^8$ cycles. Identical values of $l = 3$ and $d = 5$ are used in the
comparison.}
\begin{ruledtabular}
\begin{tabular}{l r r}
$r / \sigma$ & Error, original & Error, modified \\ \hline
1.5	& 1300000 & 5.2 \\
2.0	& 2.9	& 1.1 \\
2.5	& 2.0	& 0.96 \\
3.0	& 1.2	& 0.72 \\
3.5	& 1.3	& 0.62 \\
4.0	& 1.3	& 0.51 \\
4.5	& 1.4	& 0.41 \\
5.0	& 1.4	& 0.33 \\
5.5	& 1.4	& 0.35 \\
6.0	& 1.5	& 0.47 
\end{tabular}
\end{ruledtabular}
\label{tab:compar2}
\end{table}

One further virtue of these new angular functions is the ease with which they
can be extended to cover also the case of the interaction between non-identical
elongations and binding energies. The natural generalization of the anisotropic
functions is 
\begin{eqnarray}
\sigma(\widehat u_1, \widehat u_2, \widehat r) & = & \sigma_0 \left [1 + 
 \frac 1 2 \left \{ (l_1 - 1) |\widehat r \cdot \widehat u_1| + (l_2 - 1)
|\widehat r \cdot \widehat u_2| \right \} \right ] \\
\epsilon(\widehat u_1, \widehat u_2, \widehat r) & = & \epsilon_0 \left [1 +
\frac 1 2 \left \{ (d_1^{-1} - 1) |\widehat r \cdot \widehat u_1| + (d_2^{-1} -
1) |\widehat r \cdot \widehat u_2| \right \} \right ]
\end{eqnarray}
where subscripts $1$ and $2$ distinguish between two different types of bodies.
For a complete specification of the unlike interaction, also the corresponding
$\epsilon_0$ and $\sigma_0$ constants need be specified, something which can be
done by the standard Lorentz-Berthelot combining rules. Corresponding
generalizations for the original Gay-Berne potential have been presented by
Cleaver {\em et al.} \cite{cleaver96} and by Berardi {\em et al.}
\cite{berardi95} The latter of these generalizations covers also the case of
biaxial molecules or molecular segments.

In closing, we make a quick comparison of the computational complexity of the
modified and original potential. It is clear that eqs (\ref{eq:gay1}) and
(\ref{eq:gay2}) require a much greater number of arithmetic operations than do
eqs (\ref{eq:mod1}) and (\ref{eq:mod2}). Eqs (\ref{eq:gay1}) and
(\ref{eq:gay2}) also require the non-elementary square-root operation. In
contrast, eqs (\ref{eq:mod1}) and (\ref{eq:mod2}) require only the elementary
arithmetic operations multiplication and addition. In the author's hands, the
subroutine for the original potential requires approximately 15\% to 25\% more
computer processing time (tested on two different computers). Using fast
floating-point ({\em e. g.} less accurate) libraries is likely to diminish this
difference somewhat. In actual simulations, the speed benefit will be somewhat
less depending on system size.

\end{document}